\documentclass[a4paper]{article}
\pdfoutput=1
\usepackage{titling}
\pretitle{
\vskip .75em plus .25em minus .25em
\Huge\bfseries
}
\posttitle{
\noindent\vrule height 1.5pt width \textwidth
\vskip .75em plus .25em minus .25em
}

\usepackage[utf8]{inputenc} 
\usepackage[T1]{fontenc}    
\usepackage{hyperref}       
\usepackage{url}            
\usepackage{booktabs}       
\usepackage{amsfonts}       
\usepackage{nicefrac}       
\usepackage{microtype}      
\usepackage{epigraph}
\usepackage{array}
\usepackage{graphicx,multirow}
\usepackage{fancybox}

\usepackage[square, sort,comma, numbers]{natbib}
\setcitestyle{square}
\usepackage{geometry}
\usepackage[normalem]{ulem}	
\usepackage[toc,page]{appendix}	
\usepackage{blindtext}
\usepackage{setspace}

\usepackage{authblk}

\newcommand*\samethanks[1][\value{footnote}]{\footnotemark[#1]}

\title{Ethics of Artificial Intelligence in Surgery}
\date{}
\author[1,2]{Frank Rudzicz
    \thanks{Equal Contribution. Correspondence: [frank|raeidsaqur]@cs.toronto.edu. In Hashimoto D.A. (Ed.) \textbf{Artificial Intelligence in Surgery: A Primer for Surgical Practice}. New York: McGraw Hill. ISBN: 978-1260452730.}
}
\author[1,2]{Raeid Saqur\samethanks}
\affil[1]{Department of Computer Science, University of Toronto} 
\affil[2]{Vector Institute for Artificial Intelligence}

\begin{document}
\maketitle
\fbox{\begin{minipage}{0.95\textwidth}
    {\bf Highlights - Key Concepts:}
    \begin{enumerate}
        \item The four key principles of bio-medical ethics from surgical context.
        \item Implications of `fairness', and the taxonomy of algorithmic bias in AI system design.
        \item The shift in ethical paradigm as the degree of autonomy in AI agents evolves.
        \item The dynamic nature of ethics in AI, and the need for continuous revisions with the evolution of AI.
    \end{enumerate}
\end{minipage}}

\epigraph{\em You cannot learn to play the piano by going to concerts.}{Francis Moore [2000]}

\epigraph{\em A compass [will] point you True North from where you're standing, but it's got no advice about the swamps and deserts and chasms that you'll encounter along the way. If in pursuit of your destination, you plunge ahead, heedless of obstacles, and achieve nothing more than to sink in a swamp... What's the use of knowing True North?}{{\em Lincoln}. Dir. Stephen Spielberg. Perf. Daniel-Day Lewis, 2012}

{\setstretch{1.5}  
\section*{Prelude}\label{intro}

Surgery manifests in an intense form of practical ethics. The practice of surgery often forces unique {\em ad hoc} decisions based on contextual intricacies in the moment, which are not typically captured in broad, top-down, or committee-approved guidelines. Surgical ethics are principled, of course, but also pragmatic. They are also replete with moral contradictions and uncertainties; the introduction of novel technology into this environment can potentially increase those challenges.

A discussion about ethics is often a discussion about choice. \citep{Wall2013} defined an `ethical problem' as ``{\em when an agent must choose between mutually exclusive options, both of which either have equal elements of right and wrong, or are perceived as equally obligatory. The essential element that distinguishes an ethical problem from a tragic situation is the element of choice}.'' Moreover, choosing between options often involves identifying factors by which those options are {\em not} exactly equal, and the method one uses to weigh these factors can draw upon a set of ethical frameworks that, themselves, can be somewhat incongruous.

At their core, artificial intelligence (AI) systems -- and machine learning (ML) more specifically -- are also designed to make choices, often by categorizing some input among a set of nominal categories. In the past, the choices these systems made could only be evaluated by their correctness -- their accuracy in applying the same categorical labels that a human would to previously unseen inputs, like whether an image contains a tumour, or not. As these systems are increasingly  used in less quixotic (or more critical) scenarios, we are asking of them to make choices for which even humans struggle to find correctness \citep{Schwartz2015,Jordan2001}. Indeed, when software is more accurate than the most correct humans \citep{Esteva2017,Devlin2018}, but it is validated by labels {\em provided by} humans, the very nature of the process -- and its application into practice -- is called into question. Indeed, we may no longer consider that the human and the machine are in contrast to one another, or even that one simply utilizes the other; rather, we may consider that the surgeon and their tools are in a sense a single, hybrid, active entity. We shape our tools, and thereafter our tools shape us, as Marshall McLuhan described.


The expectations of surgery and machine learning are similar in some ways. Both surgeons and machine learning algorithms are meant to solve complex problems quickly, with a dispassionate technical skill, and neither have traditionally been defined by their bedside manner \citep{Angelos2018}. However, surgeons are regularly faced with profound ethical and moral dilemmas, often on a daily basis, as a fundamental matter of their profession, whereas establishing practical ethical frameworks in AI have only just recently begun to take shape.

In this chapter, we approximate root-cause analyses \citep{Dubois2008}, in which the factors that cause ethical dilemmas are predictable and therefore learnable through observation. \citep{Wall2013} identified broad categories of these factors in which differences (in understanding and otherwise) may lead to conflict in the surgical domain in general. These are summarized in Section \ref{sec:ethics_surgery}. Subsequently, we also explore the ethical dilemmas arising from the intertwining of AI and surgery. 

\section{Ethical aspects of surgery }\label{sec:ethics_surgery}

As issues of choice permeate ethics, it is no surprise that one has a choice of moral frameworks to apply in surgery. For instance, we may apply the {\em prima facie} duty theory \citep{emanuel1995beginning}. This is a standard widely used in biomedical ethics to provide first principles, or ground rules, for making ethical decisions in healthcare. These principles are: \textit{respect for patient autonomy, non-maleficence, beneficence, and justice}. However, these principles are not sufficient, and can themselves create conflict and ethical dilemma.  It is useful to understand these key principles to realize the inherent tension among them, and how they can give rise to ethical dilemmas in real-world scenarios. Specifically,

\begin{description}
\item[Autonomy] dictates that each medical care provider should respect the patient's choices with regards to their medical care. This principle is often put into action through informed consent, though it also means that a physician should be willing to consider alternate treatment methods if a patient rejects the original plan deemed best by the physician. However, the degree to which this principle can be upheld is often hard to clearly discern. \citep{Wall2013} pointed out that autonomy is continuously present in classical medical ethics, but in surgery it is clearly not, since while the patient is under anesthesia, shared decisions cannot be made.
\item[Non-maleficence] in bioethics is often juxtaposed with the Hippocratic oath  `to do no harm.' This requires that doctors not provide treatments or perform interventions that are needlessly harmful to patients. Surgery is somewhat unique in that to achieve some treatment goal, one has to temporarily do some harm, but that harm is typically necessary, with minimized effects.
\item[Beneficence] is the duty of physicians to maximize benefits and minimize harm to patients. In surgery, `beneficence' is a subjective measure in that the risks taken should be commensurate with the benefits received, ascertained in the context of the patients' goals and values. 
\item[Justice] stipulates that equals be treated equally and non-equals (appropriately) unequally [\citep{jonsen1982clinical}]. In health-care, justice is often described in terms of the `fairness' of the distribution of resources, including medical goods and services, or benefits and burdens of medical research. True fairness may be forever unobtainable, especially when constraints like resource scarcity force decisions that otherwise would not be necessary. Indeed, the very definition of `fairness' may be unbound. Is a utilitarian view of fairness preferable to one focused on each individual? Are systemic or institutional constraints more pressing than an individual's? Consider the case of admitting a critically ill patient to an at-capacity ICU. If available resources cannot be redistributed to accommodate the new patient, then prioritizing patients suddenly becomes necessary, and the weighting of criteria in that prioritization may be an unsolved problem. 
\end{description}

Those four key principles are not absolute, either in combination or even individually. For instance, a patient's aptitude or cognitive faculties can modulate their autonomy; consider the case of a patient with advanced dementia or other cognitive impairment -- to what extent can an understanding of the fundamental risks be truly shared between doctor and patient? To what extent is informed consent {\em truly and completely} informed when the administration of the process cannot assess for the level of information even in cognitively healthy individuals? To what extent does autonomy {\em depend} on situational awareness? 

Alternatively, \citep{Little2001} identified five moral pillars (or categories) more specific to the surgical experience, that touch on the ethical, relational, existential, and  experiential aspects, drawing on his own practice and various pieces of literature \citep{McCullough98, Frank91}. These are:

\begin{description}
\item[Rescue] Surgery involves a severe power relationship in which patients choose to surrender themselves to the surgical team \citep{Brody93}, but it is a surrender with the expectation of rescue. \citep{Little2001} suggested that this first pillar of surgical ethics must be acknowledged and negotiated between patient and surgeon, and can serve as the basis for a shared understanding between them. If the rescuer is a human, the patient may have certain expectations as to the surgeon's emotional involvement, and their subsequent urgency in the rescue relationship. If the rescuer is increasingly augmented by automation, where no emotional urgency of rescue may be expected, this pillar may begin to crumble.
\item[Proximity] Surgery is at the extreme of the `medical gaze' in which the surgeon has a physical proximity to the patient, and their inner workings, that even the patient themselves cannot typically observe. Proximity to the patient is not only physical, however, but emotional and personal, and often the surgeon will install limits on this type of proximity as a means of self-preservation, to avoid the penalties of closeness or the pains of failure \citep{Little2001}. Unemotional aspects of automation clearly do not require layers of emotional protection, but they introduce other facets to consider. For example, if a purely data-driven analysis of a patient or case reveals patterns that are too intricate or complex for the typical doctor-patient relationship to capture, to what extent can they be acted upon? If they are acted upon, how can they be rationalized or explained? This is discussed in Section \ref{sec:xai}.
\item[Ordeal] The ordeal of surgery requires that a patient concede autonomy, face risk and possibly mortality, suffer physical pain and discomfort, and disrupt the flow of their lives. When the benefits of rescue override the trauma of the ordeal, those aspects are usually tolerated. Sometimes, the depth of the ordeal is not fully grasped prior to a surgery, and negative outcomes predictable to the surgeon but not the patient may raise conflict after the fact. \citep{Cassel1982} suggested that relieving suffering and curing disease are in fact twin obligations, and that a technically adequate procedure may not be a sufficient intervention if the nature of the suffering is not also considered. 
\item[Aftermath] Physical and psychological scars persist after surgery, and surgical ethics must accommodate both objective and subjective aspects of these experiences. \citep{Little2001} does not make it explicit, but the scope of the aftermath may depend on intra-operative factors, and adverse events that may be common nevertheless can directly magnify the aftermath, including risks of re-admission, death, lifelong injury, or other postoperative complications. For these reasons, it will be crucial to minimize the scope and profundity of the aftermath by using data-driven analytics, including AI, in the prediction and prevention of adverse events \citep{Jung2018,Goldenberg2017}.
\item[Presence] Since a patient surrenders themselves to the surgical team, they will typically expect or desire a certain degree of access to that team (at least, to the extent that the access is initiated by them). Personal attributes of charisma, confidence, and empathy may not be as important as the stamina, cognitive abilities, and mere time that a surgeon can present to their patients. Being available gives a sense of comfort and protection. From this perspective, it does not matter if an AI remains less personable than a surgeon -- it may be {\em more} relevant that the information processed by automated processes is continuously available and interpretable to a patient.
\end{description} 

Rather than broad principles, \citep{Wall2012} outlined a set of questions with definite possible responses, clustered by category, that can be used to identify the root cause or causes of an ethical dilemma, as recreated in Table \ref{tab:WallQuestions}. These questions can be adapted to the scenario, as flexibility is often necessary in clinical contexts, but the point is to have some objective set of fundamental variables from which a computation of ethics is to be made.

\begin{table}[h!b]
    \centering
    \begin{tabular}{c|l}
        \hline
         \multirow{3}{*}{\rotatebox{00}{Stakeholders}} & Who are the primary stakeholders?  \\
         & Are there additional medical providers who should be involved?\\
         & Does the patient have additional people who should be involved?\\
         \hline
          \multirow{5}{*}{\rotatebox{00}{Facts}} & What is the diagnosis?\\
          &What are the options for intervention?\\
          & What are the expected outcomes for each intervention (including no intervention)?\\
          & What does each of the stakeholders understand about the medical facts?\\
          & Are there additional facts relevant to the case?\\
         \hline
          \multirow{4}{*}{\rotatebox{00}{Goals and values}} & What are the goals of each of the primary stakeholders?\\
          & Are there any conflicts among the various goals?\\
          & What are the values of the primary stakeholders?\\
          & Are there any conflicts among the identified values?\\
         \hline
          \multirow{5}{*}{\rotatebox{00}{Norms}} & What are the relevant ethical norms?\\
          & What are the relevant legal norms?\\
          & What are the relevant institutional norms?\\
          & What are the relevant professional norms?\\
          & Are there conflicts among the identified norms?\\
         \hline
    \end{tabular}
    \caption{Questions to ask prior to the decision-making process, according to \citep{Wall2012}.}
    \label{tab:WallQuestions}
\end{table}

In contrast to the more traditional {\em prima facie} duty theory, or the moral pillars of \citep{Little2001}, each of which broadly outline qualitative principles, \citep{Wall2012}'s framework of identifying atomic facts of a case may be more suited to a quantitative {\em computation} of ethics. That is, given that we have a set of objective facts, then one ought to be able to compute an objective ethical outcome given an objective transfer function with sufficient parameters. In other words, if one has a set of variables $X=\{ x_1, x_2, ..., x_N \}$ that answer, for example, the questions in Table \ref{tab:WallQuestions}, we may seek a mathematical function $f(\cdot)$ that takes $X$ as its input and outputs some numerical score $S=f(X)$ of the {\em utility} for one decision over all others. Whether ethics can ever be truly {\em calculated} in the medical context should be open to debate, but it is worth noting that this sort of cost-benefit calculation is precisely what artificial agents perform when deciding on what action to take, in AI's subfield of `reinforcement learning' \citep{Abel2016}.

\section{Ethical aspects of using AI in surgery}\label{ethics_surgery}


In the preceding section, we presented multifarious frameworks for surgical ethics, discussed the {\em prima facie} duties of human surgeons as full ethical agents animated by certain key principles \citep{emanuel1995beginning}, and outlined questions to ask when analyzing ethical problems. 


Machine learning in practice includes all of the risks typical of software generally -- malware and hackers, software bugs and lack of dependability, and disparities and inequities inherent in the society in which the software is used. In surgery, applications of AI are largely confined to the agents performing specific and defined tasks initiated and controlled entirely by human surgeons or clinicians, although recent advances such as the Smart Tissue Autonomous Robot (STAR) have begun to perform {\em in vivo} and {\em ex vivo} surgical tasks more effectively than humans, albeit in quite specific and controlled conditions \citep{Panesar2019}. AI may also be used as a clinical decision support system prior to surgery or other acute care \citep{Lynn2019}. In such circumstances, the ethical paradigm may be fully enveloped using de-ontological (i.e., \textit{defined duty}) ethics. In this framework, an AI system is an \textit{implicit ethical agent}, and the burden of responsibility of machine behaviour falls on the human designers and developers of the machine agent or AI system. Thus, we focus on ethical issues faced by human designers in developing AI systems in the current context. 

\subsection{Privacy of (mostly patient) data}\label{privacy}

The decision-making capability of an AI system emanates from an underlying mapping  between relevant inputs and an output decision. 
Regardless of the specific learning mechanism, training a production-grade AI system for surgical applications usually requires a non-trivial amount of sensitive patient data. The \textit{privacy} of such patient data is therefore an automatic-choice topic in many discussions concerning ethics of AI in surgery, because it can be argued that personal medical data are the most sensitive kind of user data available. Hence, the safe-guarding of users' medical data is of the utmost importance and a primary concern when applying AI in surgery. In this way, the data-hungry algorithms of modern machine learning are traditionally described as being at-odds with modern ethical approaches to patient data, but this need not be the case \citep{Rohringer2019}, as there are various emerging tools, including {\em federated learning}, which can keep an individual's data decentralized and while still approximating the performance of a system with access to a more complete data set \citep{Bonawitz2019}. An alternative approach to keeping patient data private is {\em differential privacy} in which the outputs of a statistical system are provably uninformative as to whether an individual's data are included \citep{Wu2016}. This is accomplished, essentially, by judiciously adding noise in a way that does not fundamentally hinder the accuracy of the system, but makes an adversarial attack nearly impossible, even with auxiliary information \citep{Dwork2008}. Other algorithmic protections to patient privacy are constantly being developed.

Keeping pace with the development of AI systems, there has been a recent surge in regulatory measures addressing data privacy. A prominent example is the European General Data Protection Regulation (GDPR), which not only stipulates stringent data privacy measures for direct patient care, but also for research \citep{mccall2018does}. The American Health Insurance Portability and Accountability Act (HIPAA) provides similar stipulations.  These measures are somewhat limited by their jurisdictional  interpretations. Juxtaposing GDPR with HIPAA, we see that although these regulations may have similar overarching intents, their scopes are different. HIPAA focuses on healthcare data from patient records, but does not cover alternate sources of health data generated outside of covered entities, such as life insurance companies or applications on smart devices (e.g., fitness bands, smart-watches) \citep{cohen2018hipaa}. These frameworks also offer incomplete models of ethics. For example, the duty of patient autonomy and their right to `informed consent' alludes to a `right to explanation' of all decisions made by automated or AI systems \citep{Goodman2016}. The right to explanation was widely expected to be legally mandated by GDPR once enforced \citep{wachter2017right}; however, in reality, GDPR only mandates a `\textit{right to be informed}' -- i.e., patients could receive meaningful information (Articles 13-15), however limited, about the logic involved, as well as the significance and envisaged consequences of automated decision support systems.  In this area, GDPR appears to lack precise language to explicitly state well-defined rights and safeguards against AI-based automated decision making and, therefore, may limit its ineffectiveness. However, given the complexity and state of explainability in modern AI models, it is currently infeasible to stipulate and enforce the full scope of a `right to explanainbility' for AI systems, which we discuss in more detail in Section \ref{sec:xai}. Indeed, given the novelty of machine learning generally, there remains significant confusion over which laws and regulations might apply \citep{Cortez2018}.

Although the focus of the discussion on privacy has typically been on patient data, there are also, naturally, concerns about the privacy of the surgeon and their teams, as well. As AI is increasingly being used by tools that assess or analyze aspects of the surgery itself \citep{Goldenberg2017}, concerns may be raised around increased legal risk, or even changes to behaviour (and therefore performance) given that one knows they're being recorded. Jurisprudence reveals a few principles in Western law applicable to recording surgical procedures, touching on the same issues of privacy as for the patient, with the addition of the importance of professional secrecy \citep{Henken2012}. 

\subsection{Fairness and algorithmic bias in AI} 
The issue of fairness in using AI systems primarily concerns the inadvertent \textit{algorithmic biases} and {\em statistical biases} inherent in the design and functioning of an AI system. Due to the magnitude and severity of the consequences that these biases may engender -- including in social, legal, and medical domains -- the topic of `fairness' ranks as one of the top recurring themes in machine medical ethics and AI ethics in general. Biased AI systems can have various deleterious effects in multiple domains, and the types of algorithmic biases can impact the surgical cases presented in the other chapters of this book. 

Decision-support systems powered by AI -- designed with controls for fairness -- may be used to augment human judgment and reduce both conscious and unconscious biases \citep{anderson2007machine}. In contrast, unintended statistical biases caused by differences in sampling (or insufficient controls such as data regularization) may reflect and amplify existing cultural prejudices and inequalities \citep{sweeney2013discrimination}. This is reflected in a litany of research confirming racial \citep{barr2015google, garcia2016racist, bass2017researchers,sweeney2013discrimination} and gender bias \citep{bolukbasi2016man} emanating from insufficiently examined autonomous decision support systems. For example, comprehensive studies by \citep{kehl2017algorithms} and \citep{flores2016false} showed systemic racial bias in algorithms used to predict recidivism, with implications for the US Criminal Justice system. In natural language computing, models of language derived from general texts using neural networks have learned a host of gender biases with which we may not be comfortable (e.g., that a man is to a woman as a programmer is to a homemaker, or as a surgeon is to a nurse \citep{bolukbasi2016man}. While progress can be made in minimizing biases against identifiable groups, since machine learning is meant to uncover complex non-linear relationships in the data, it is quite possible that groups (or procedures, or treatments, etc) that are less well identified in the societal zeitgeist will be isolated by these pattern-finding algorithms, and nevertheless not be detected by the system's users \citep{Char2018}. 

\citep{gianfrancesco2018potential} and \citep{ Vayena2018} focused their discussions predominantly on `\textit{training data bias}' -- which concerns the quality and representativeness of data used to train machine learning and AI systems. 
A taxonomy of biases in machine learning may be dichotomized into those originating from either 1) technical or computational sources, or 2) inappropriate use or deployment of algorithms and autonomous systems. Statistical bias in training data falls under the first category. Adhering to the iconic aphorism -- `\textit{garbage in, garbage out}' -- bias in the input will result in a biased model. For example, existing medical datasets have had much higher ratios of adult males of Caucasian origin (i.e., an over-representation bias) than exists in the actual population \citep{landry2018lack}. A lack of diversity in sampling manifests in biased data and, without special controls in place, therefore results in biased models that may not behave as expected for under-represented groups \citep{Adamson2018}. Similarly, EMR data can suffer from missing samples and incorrect labels \citep{gianfrancesco2018potential}, with unintended downstream effects.  
The sub-optimal performance for underrepresented social groups creates an ethical bottleneck.

Other sources of technical or computational bias include algorithmic \textit{focus} and \textit{processing} bias. The \textit{focus bias} emanates from the differential usage of information in training AI systems. For example, developers may  deliberately include or exclude certain features (i.e., types of inputs) when training a model, thereby causing it to deviate from the statistical standard if those attributes have a strong main effect on the outcome, or interaction effects with other variables. \textit{Algorithmic processing} bias occurs when the algorithm itself is biased, as in the use of statistically biased estimators, which may result in a significant reduction of model variance on small sample sizes (i.e., the bias-variance tradeoff \citep{geman1992neural}). Thus, developers may embrace algorithmic processing as a bias source in order to mitigate or compensate for other types of biases \citep{danks2017algorithmic}. 

The potential effects of biases emanating from technical and computational sources,  of AI in surgery, could have direct effects on patient safety and system integrity. For example, training data bias could dramatically impact a preoperative risk stratification prior to surgery. Under-representation of demographic clusters may also cause inaccurate risk assessments and thereby AI-driven decisions, such as which patient is treated first or offered resources in the ICU post surgery, will be flawed.

\textit{Transfer of context} bias, and \textit{interpretation} bias are two remaining sources of algorithmic biases that constitute inappropriate use or deployment of algorithms and AI systems \citep{danks2017algorithmic}. The transfer of context bias is introduced when an algorithm or trained model is employed outside the context of its training. An unwarranted application or extension of an AI system outside of its intended contexts will not necessarily perform according to appropriate statistical, moral, or legal standards. For example, a machine learning model trained only with simulated surgical data being deployed to analyze real-life patient surgeries would not be appropriate. Transfer context bias may also arise when translating a model from a research hospital directly to a rural clinic. Once again, the machine learning community has identified the challenge and has begun to tackle this challenge, as in off-policy learning \citep{Rakelly2019}, but as these challenges become more subtle, so too do their solutions and hence it is crucial that all stakeholders are kept up-to-date about the state-of-the-art.

Consider the case of automated intra-operative video analyses during surgery. Contextual bias {\em may} occur if one ignores what may appear to be trivial caveats (e.g., whether the operating surgeon is right or left-handed), or if one presumes that the same level of resources are available across operating rooms and hospitals. Methods to mitigate against this include ensuring variance in the data set, being sensitive to overfitting on training data, and having humans-in-the-loop to examine new data as it's deployed, for starters.

Lastly, \textit{interpretation bias} concerns the misinterpretation of an algorithm's output by either a human user or by the broader system within which the algorithm functions. Developers are rarely able to fully specify all possible scenarios in which their algorithms or models are to be used. For example, biased judgments about causal structure or strength can be easily misused in biased ways by AI systems if not careful accounted. 

\subsection{Transparency and explainable AI (XAI)}\label{sec:xai}
The best performing modern AI models predominantly use deep neural networks, which can be abstracted and thought of as black-box, non-interpretable function approximators. Thus, even though such models can be used to achieve high prediction accuracies, one cannot {\em a priori} coherently explain or attribute the influence of particular features of input on the resulting predictions in a causative manner, nor necessarily describe the model itself in human-graspable terms.  At least, not by default.

Consider a simplified binary classifier that takes in patient data $x$ and outputs whether the patient has cancer or not, $y$ (i.e., $y=+1$ for cancer, $y=-1$ otherwise). The input data here could be a patient's biographical information, cell culture images, etc. In order to train the model, input-label pairs \((x, y) \in \mathcal{X} \times\{ \pm 1\}\) are sampled from a data distribution $D$; the goal is to learn a classifier \(C : \mathcal{X} \rightarrow\{ \pm 1\}\) which predicts a label $y'$, corresponding to a particular new input $x'$. We can define a \textit{feature} to be a function mapping from the input space $\mathcal{X}$ to the real numbers, with the set of all features being \(\mathcal{F}=\{f : \mathcal{X} \rightarrow \mathbb{R}\}\). In this case, the patient's weight, height, age, BMI, and so on, could be considered features, represented by normalized real numbers. For an arbitrary `black box' deep neural network, the precise influence that function $\mathcal{F}$ has in the mapping of $\mathcal{X}$ to a binary prediction $y \in \mathcal{R}$ will be lost in the layers upon layers of artificial neurons or, at the very best, the mapping will be so complex as to be uninterpretable. Thus, even if an ML system is able to predict  cancer with very high accuracy, one could not infer or attribute which features \(f \in \mathcal{F}\) caused this prediction, and hence one may find it difficult to take interventional action, if the causes of the diagnosis -- especially the modifiable ones -- are unknown. 

If a system's output cannot be interpreted, it invalidates its use in a broad swath of domains when formulating \textit{health care policies}. In a surgery setting, if more diagnostic and therapeutic interventions become based on AI, the autonomy of patients in decisional processes may also be undermined. Already, an increasing reliance on automated decision-making tools has reduced the opportunity of meaningful dialogue between the healthcare provider and patient \citep{Vayenaid2018}. However, the current ethical debate of whether (or how) the details of an AI system should be disclosed to patients is still in its infancy. A bigger ongoing concern is whether such black-box algorithms, whose self-learned rules may be too complex to reconstruct and explain, should be used in medicine \citep{price2014black}. Although some have called for a duty to transparency to dispel the opacity such systems \citep{wachter2017right}, others have justified limited requirements -- for example, the \textit{right to be informed}, instead of the \textit{right to explanation}, suffices to adequately protect the morally relevant interests of patients when AI algorithms are used to provide them with care \citep{selbst2017meaningful}.

If machine learning is in its infancy, then the subfield tasked with making its inner workings explainable is so embryonic that even its terminology  has yet to recognizably form \citep{Adadi2018}. However, several fundamental properties of explainability have started to emerge. Among these, \citep{Lipton2016} argued that machine learning should be {\em simultaneous} (its behaviour should be graspable holistically), {\em decomposable} (its components or inputs should also be understandagble, and {\em algorithmically transparent} (the shape of the solution, and the method to get there, should be somewhat intuitive). Moreover, explanations can be textual, visualized, local to a data point under consideration, or in terms of other examples -- prototypical or critical.

The benefits of adding explainability and transparency to the decision-making process is illustrated by a case study in anesthesiology \citep{lundberg2018explainable}, where a clinical decision support system  was used to predict hypoxaemia in patients in real-time during surgeries. The system was trained on EMR data combined with minute-by-minute vitals data over 50,000 surgeries and helped to increase the rate of anticipating hypoxaemia from 15\% to 30\% during surgeries. However, the system not only made predictions, but also provided real-time interpretable risks in terms of an automatically selected subset of contributing factors. This allowed for easier early intervention because the results could be associated with modifiable factors. Additionally, this system can help improve the clinical understanding of hypoxaemia risk during anaesthesia care by providing general insights in the exact changes in risk induced by patient characteristics or surgical procedure \citep{Gordon2019}.

\section{Responding to challenges, duties of human designers, and institutional interventions}

Monitoring the integration of AI in surgery will involve every healthcare worker it affects. In particular the Surgeon-in-Chief, in consultation with a committee of stakeholders, can embody the institutional memory and authority to ensure that progress is managed effectively and safely \citep{Das2019}. In the current context, the responsibility of a machine's behaviour apparently falls on its human designers and developers. \citep{Stahl2016a} argued that an effective way for handling contingencies is to minimize the gap between AI researchers and the different stakeholders (or the subsequent users) of the systems they create. They propose the `Responsible Research and Innovation' (RRI) framework including ethical considerations built-in to research, i.e., at a nascent stage of technical development. Clearly, part of this will involve a more effective dialogue between developers and stakeholders, including around potential ethical contingencies. 

AI and ML are often referred to as the source of risks when discussing their use in healthcare, but the contrapositive is rarely discussed -- what are the ethical risks of {\em not} integrating these into practice, especially in situations where their effects can only be positive? What are the risks of maintaining the {\em status quo}, which has been so replete with cognitive error?

Innovation in surgery is a process rather than an event, and is commonly the result of creative attempts to solve individual problems \citep{Angelos2014}. The process will help filter good ideas from bad ones through multiple layers of scrutiny. However, too many filters may be prohibitively dense, and keep the best novel ideas from impacting care in the short term. In traditional thyroidectomy, the risks of permanent hypoparathyroidism and recurrent laryngeal nerve injury are only 1\%-2\%. If a method guided by AI can reduce those risks by a half, it may take thousands of procedures with the new method to observe any statistically significant change, which an individual surgeon may never observe, at least not in a short time frame. However, a repository of AI-based analytics, aggregating these thousands of cases from hundreds of sites, would be able to discern and communicate those significant patterns. This is the promise of scale that AI delivers.

However, scale itself can provide a subtle but concrete mode of failure, unique to AI. \citep{Amodei2016} described several problems in AI safety that often evade discussion, including:

\begin{description}
\item[Avoiding negative side effects] involves placing bounds on the negative consequences, or on impact generally, of an AI system in the pursuit of its goals. Since AI systems will typically be used in environments not identical to those in which it was trained, it is important for such systems to identify those bounds itself, automatically.
\item[Avoiding reward hacking] means setting goals that do not have `short cuts' or hidden paths to optimization that might be unexpected. For example, a system in the same vein as \citep{lundberg2018explainable}'s, designed to not only predict but {\em avoid} adverse events, may incur a penalty for every such detected event and, if not carefully controlled, may learn to ignore such detections to avoid cost.
\item[Scalable oversight] includes questions of how an AI system can explicitly evaluate its performance, especially in the long-tail of the many potential atypical or rare cases, when the size of the data grows beyond human intervention. 
\item[Robustness to distributional shift] ensures that a system behaves robustly, or can generalize its behaviour, to new data that is fundamentally different than the data on which it was trained. For example, a system trained in a large, modern, and expensively outfitted integrated OR should be able to maintain a level of performance even in smaller, more modestly outfitted environments.
\end{description}

These aspects can manifest in subtle ways, often eluding human scrutiny (which further emphasizes the need for explainable systems). In the literature, case studies are often described by contrasting extreme but recognizable aspects. While this can help to familiarize one's self with the types of frameworks traditionally used in ethics (see Section \ref{sec:ethics_surgery}), real-life ethical dilemmas are often not so extreme, and occur at the boundaries of more subtly differing aspects. Due to the domain-specific subtleties and complexities of AI systems, it's imperative to have domain and industry-specific guiding frameworks for applications of AI. 
The continued technological advancement in AI will sow rapid increases in the breadths and depths of their duties. Extrapolating from the progress curve, we can predict that machines will become more \textit{autonomous}.  
We observe a concomitantly accelerating emergence of frameworks for algorithmic fairness and impact assessment by governments and institutional authorities, including the Government of Canada's Algorithmic Impact Assessment \citep{Reisman2018} and the HART framework in the UK \citep{Oswald2018}. 
Although these (and similar) frameworks have common overarching intentions, due to the unprecedented pace of innovation, it will remain difficult to produce complete frameworks that can provide total guidance for dispelling ethical challenges, and thereby ensuring ethical AI systems. Some recent frameworks have attempted to occupy the niche between healthcare and AI, including a notable framework by \citep{Luxton2014}, who reiterates that theoretical approaches and expert knowledge must be current (which is an ever increasing challenge itself) and evidence-supported. Among his considerations for ethical codes, guidelines, and uses of AI systems in healthcare are:
\begin{itemize}
\setlength \itemsep {0em}
    \item Disclosing the services of an AI system, and their limits, to patients.
    \item Ensuring education of an AI system's users around capabilities, scope, and limitations, and for these to be communicatable during informed consent.
    \item Requiring human supervision and monitoring for adverse outcomes and risks (although this of course presumes that human monitoring and intervention would not in themselves cause error and risk).
    \item Following applicable privacy laws and best practices.
    \item Ensuring that the AI system follows established clinical best practices.
    \item Ensuring that the AI is used within a continuity of care sensitive to the emotional nature of patient interactions.
    \item Providing a mechanism for patient feedback and queries.
    \item Identifying specifications for use, including limits on autonomy.
    \item Testing safety and ethical decision-making in diverse situations.
    \item Including data logs and audit trails.
    \item Considering cultural sensitivity and diversity in designing AI systems.
\end{itemize}

\citep{Luxton2014} also nevertheless emphasizes that ``{\em formal ethical principles can never be substituted for an active, deliberative, and creative approach}''. In this way, AI development will need to be guided by an {\em ad hoc} interpretation of principles in the same way that surgeons are merely guided by the principles we discussed before, such as the {\em prima facie} theory, or \citep{Little2001}'s `moral pillars'.

The rise in autonomy necessitates an increased focus on the ethical horizon that we need to scrutinize. The discussion may soon transcend from deontological (defined duties) ethics to teleological (consequentialist) ethics. In surgery, some questions that may arise regarding  AI may include \citep{bendel2015surgical}: what limits do we place on AI autonomy? Can a surgical robot refuse to perform a surgery? How do we attribute the responsibilities of machine behaviour in this scenario, i.e., who is responsible if the machine performs poorly? Does a robot surgeon support or compete with physicians and their assistants? We intentionally refrain from elaborating and exploring these questions; a prescriptive plan for future AI systems would be, in some sense, like making life choices by gazing into a crystal ball. However, it will be important to remember 1) that the nature of ethical challenges of AI in surgery will remain dynamic for some time, due to the evolving and constantly shifting technological capabilities, and 2) increasing AI autonomy will drastically expand the ethical paradigm, and the challenges that come with it.

Ethical considerations of applying AI in healthcare generally often involve adhering to data protection and privacy requirements, fairness (broadly construed) across data sourcing, development, and deployment, and the maintenance of certain standards for transparency \citep{Vayena2018}. Like ethical decision-making in current practice, machine learning will not be effective if it is merely designed carefully by committee -- it requires exposure to the real world. 




} \label{setstrech_end}
\bibliographystyle{plain}
\bibliography{main}

\end{document}